# Concept of multichannel spin-resolving electron analyzer based on Mott scattering

Vladimir N. Strocov,[1] Vladimir N. Petrov,[2] and J. Hugo Dil[1,3]

[1] Swiss Light Source, Paul Scherrer Institute, CH-5232 Villigen-PSI, Switzerland

[2] St. Petersburg Polytechnical University, Polytechnicheskaya Str. 29, RU-195251 St Petersburg, Russia

[3] Institut de Physique de la Matière Condensée, École Polytechnique Fédérale de Lausanne, CH-1015 Lausanne, Switzerland

Concept of a multichannel electron spin detector based on optical imaging principles and Mott scattering (iMott) is presented. A multichannel electron image produced by standard angle-resolving (photo) electron analyzer or microscope is re-imaged by an electrostatic lens at an accelerating voltage of 40 keV onto the Au target. Quasi-elastic electrons bearing spin asymmetry of the Mott scattering are imaged by magnetic lenses onto position-sensitive electron CCDs whose differential signals yield the multichannel spin asymmetry image. Fundamental advantages of this concept include acceptance of inherently divergent electron sources from the electron analyzer or microscope focal plane as well as small aberrations achieved by virtue of high accelerating voltages, as demonstrated by extensive ray-tracing analysis. The efficiency gain compared to the single-channel Mott detector can be a factor of more than $10^4$ which opens new prospects of spin-resolved spectroscopies in application not only to standard bulk and surface systems (Rashba effect, topological insulators, etc.) but also to buried heterostructures. The simultaneous spin detection and fast CCD readout enable efficient use of the iMott detectors at X-ray FEL facilities.

**Keywords:** Electron spin, Mott scattering, electron optics, angle-resolved photoelectron spectroscopy and microscopy, X-ray free-electron lasers

# 1. INTRODUCTION

The spin of the electron plays a crucial role in many physical phenomena, ranging from the obvious example of magnetism, via novel materials for spintronics applications, to high temperature superconductivity. The direct detection of the spin has therefore played an important role in the understanding of processes such as giant magneto-resistance (GMR),[1,2] the formation of ferromagnetic domains,[3] the Rashba effect[4,5] and topological insulators.[6,7] Starting from the classical experiments of Schönhense,[8] it has been realized that any angle-resolved photoelectron spectroscopy (ARPES) experiment yields in fact spin-polarized intensities. The most direct access to electronic structure of crystalline solids resolved in electron spin and momentum is therefore delivered by spin-resolved ARPES (SARPES), for a recent review of the field see Ref.9. In the aforementioned examples the focus was on the detection of the spin of the valence and core level states, but it has also been realised that the spin polarization as induced by the excitation process can yield a rich variety of information.[9] Such measurements typically require the possibility to vary the light polarization and/or energy, but then provide a rich variety of information such as magnetic circular dichroism above the Curie temperature,[10,11] the distinction between singlet and triplet spin states,[12] and the study of unconventional superconductors.[13] The appeal of these possibilities has led to a revival in the interest of spin detection in combination with spectroscopic techniques and the construction of new experimental setups primarily using synchrotron radiation sources with their high brilliance and tuneable photon energy $h\nu$.

On the technical side, the general evolution trend of the electron as well as X-ray spectroscopic instrumentation can be characterized by moving from the use of slits for selecting a certain energy channel to the use of dispersion and imaging principles for multichannel acquisition in an extended region of phase space. For ARPES, in particular, the first major hallmark of this development vector has been the use of the dispersing and focusing properties of the hemispherical capacitor to produce a multichannel one-dimensional (1D) image of electron intensity $I_{ARPES}$ over certain region of kinetic energies $E_k$. The second hallmark has been an invention of electrostatic lenses which could image the emission angles $\theta$ onto the entrance slit of the hemispherical capacitor to produce in its exit plane a multichannel two-dimensional (2D) image of $I_{ARPES}(E_k,\theta)$ over certain region of $E_k$ and $\theta$.[14,15] Finally, the angle-resolved time-of-flight (ARTOF) analyzers[16] allow multichannel acquisition of three-dimensional images of $I_{ARPES}(E_k,\theta_x,\theta_y)$ as a function of $E_k$ and two orthogonal emission angles $\theta_x$ and $\theta_y$. Similar development in the resonant inelastic X-ray scattering (RIXS) has been marked by the spectrometers based on the Rowland-circle[17] and variable-line-spacing[18,19] spherical gratings whose combined dispersion and focusing actions produced a multichannel 1D-image of scattered X-ray intensity $I_{RIXS}$ over certain region of scattered energies $h\nu_{out}$. Finally, the concept of an $hv^2$-spectrometer[20,21] has been advanced where simultaneous imaging action in the orthogonal plane produces a full multichannel 2D-image $I_{RIXS}(hv_{in},hv_{out})$ as a function of incoming $hv_{in}$ and scattered $hv_{out}$ energies. The overall efficiency gain due to the multichannel detection in the 2D-case compared to the single-channel detection can measure for various instruments up to 4 and more orders of magnitude.

Although new exciting schemes of more efficient spin detection have recently been established[22] the main technical difficulty of SARPES remains a dramatic intensity loss. This is characterized by the figure of merit (FOM) $F = S^2 \frac{I}{I_0}$ where $\frac{I}{I_0}$ is the scattered to incident intensity ratio, and $S$ the asymmetry (also called Sherman) function defined as the measured intensity asymmetry $A = \frac{I^+ - I^-}{I^+ + I^-}$ between two opposite polarization directions divided by the electron polarization $P$. With small FOM values less than $10^{-2}$ for one spin projection, high quality spin-resolved measurements are primarily possible only at synchrotron radiation facilities or using high power lasers, and then still only within a limited range of experimental parameters and/or using measurement times of many hours or even days. This has inhibited the expansion of spin detection to other techniques which may be inefficient in itself because of a lower cross-section, or because they rely on detection of the time structure or because they sample only a small amount of matter. Obviously, SARPES and other spin-resolving techniques appear in most severe need of the multichannel detection. Surprisingly, only most recently have such instruments come into play, starting from photoemission microscopes with their natural multichannel detection. The spin resolution was achieved here using spin-polarized reflectivity of a collimated low-energy (below 100 eV) electron beam from W(100) or (Au-passivated) Ir(100) crystals working as imaging spin filters.[23,24,25] Furthermore, installation of such spin filter behind the standard hemispherical analyzer (HSA) has for the first time enabled spin-resolved energy- and angle-multichannel ARPES measurements[26] although suffering from the inherent divergence of electron trajectories in HSA (see the discussion later).

The Mott spin detectors (for entries see Ref.27,28,29) are based on quasielastic scattering of high energy electrons (about 40 keV) from a target of a high-Z material such as Au. As a spin selective process, the Mott scattering is characterized a FOM value about $6\times10^{-4}$ per one spin projection[27,28] measured by two opposite electron detectors working simultaneously. Other spin selective processes such as the low-energy electron reflectivity from W(100) or Ir(100)[23,24,25,26] or exchange scattering from the $Fe_2O_3$ surface[30,31] may certainly have larger FOM values, in the latter case reaching $9.5\times10^{-3}$ per one spin projection[30] acquired in two successive measurements under re-magnetization of the target or sample. However, the Mott detectors, brought to equal efficiency in all channels by proper calibration,[27,28] allow simultaneous measurements of two spin projections, which doubles their effective FOM to an already comparable value of $1.2\times10^{-3}$.

From a practical point of view, the FOM analysis alone ignores other aspects of the spin-resolved experiment. First, the simultaneous differential measurements render the Mott detectors immune to the statistical fluctuations brought, for example, by those of the excitation source such as the synchrotron beam or ripple of voltages on the electron optics.[32] Furthermore, the long-term stability and reliability of the Mott detector are just as important for obtaining reliable data because of potential degradation of the sample during the acquisition time and drifts in the electron optics. This can be illustrated, for example, by recent studies of the topological Kondo insulator $SmB_6$ where the spin-polarized surface states have a low spectral intensity and are located in a 50 meV band gap. Although attempt were made using a detector with a higher FOM[33] only the Mott detectors provided enough long term stability to explore this spin texture.[34] The reliability of the Mott detector is expressed, in particular, in the fact that the high energies make it insensitive to the surface quality of the target, and the detector will function in the same way every time it is switched on and for a long period of time. This is critical at large scale research facilities such as synchrotrons. A further major advantage of the Mott detectors lies in the fact that the spin contrast is obtained without repeating the measurement under different conditions such as flipping the spin of the incident electron beam in the electron excitation experiments or re-magnetization of the target or sample. This simplifies studies of non-magnetic samples and, most important, enables experiments under conditions where an exact repetition is fundamentally impossible, such as studies on quickly degrading samples or using free electron laser (FEL) pulses. We note that the above practical advantages are relevant only to the classical-type Mott detector where the electrons scattered off the target move in field-free space. In the retarding- or Rice-type Mott detectors the scattered electrons move in retarding potential (for comparative tests of various Mott detectors see Ref. 32). Such detectors are actually highly sensitive to parameters of the incident beam, which results in unstable measurements of the polarization asymmetries.

The most important advantage of the Mott detectors appears however in the multichannel detection perspective. By virtue of high electron energies, the Liouville-Helmholtz theorem works in these detectors completely in advantage of the electron optics, which delivers then minimal aberrations at large angle and energy acceptance, as will be explained below. Here, we present a concept of angle- and energy-multiplexing spin-resolving electron analyzer which combines electron optics based on the imaging principles with the detector based on the Mott scattering. This "imaging Mott" detector will be nicknamed "iMott".

## 2. OPERATION PRINCIPLE

We will sketch the iMott concept as tailored to the standard angle-resolving HSA. Fig. 1 (*a*) shows the iMott detector attached to HSA. The latter creates in its exit (focal) plane an image of electrons dispersed in the X and Y coordinates corresponding to $E_k$ and $\theta$. Fig. 1 (*b*) shows a blow-up of the iMott itself. The electrostatic lens (*EL*) operating at high voltage accelerates the electrons from HSA focal plane and images them onto the polycrystalline Au target to create there a demagnified image again stretching along the $E_k$ and $\theta$-directions. The electrons quasi-elastically scattered from the Au target and bearing spin asymmetry of the Mott scattering are then imaged by four magnetic lenses (*ML*s) (chosen instead of electrostatic ones because of smaller aberrations with extended sources) onto the energy-selective and position-sensitive detectors (in our case implemented as electron-sensitive CCD – *eCCD* –detectors) to create there images of scattered intensity stretching along the $E_k$ and $\theta$-directions (distortions of these images by the magnetic field is corrected on the post-processing stage). The differential images between the opposite *eCCD*s immediately yield a multichannel image of the $A(E,\theta)$ spin asymmetry in the *E*- and $\theta$-coordinates. In addition, another two MLs and eCCDs installed in the orthogonal scattering plane will simultaneously measure the orthogonal spin component. As we will see below, the overall efficiency gain granted by our multichannel concept can be well above $10^4$.

## 3. RAY-TRACING ANALYSIS

We will now illustrate the feasibility of the iMott concept with ray-tracing simulations. We restricted ourselves to a compact design of the instrument with a radius of the Mott hemisphere $R_M$=100mm, allowing its seamless use with virtually any experimental setup. Without restrictions on generality, we will tailor our analysis to the spatial and angular characteristics of the electron trajectories in the commercial analyzer PHOIBOS-150 having a hemisphere radius of 150 mm, which operates in the Medium Angle Acceptance mode with an entrance slit of 200 μm and pass energy $E_p$ of 40 eV.[35] The energy resolution $\Delta E$ and angle resolution $\Delta\theta$ are in this case 36 meV and 0.07° FWHM, respectively, with the latter achieved with the source size below 100 μm typical of the synchrotron radiation sources. For the simulations we select a beam of incident electrons comprising 5 values of $E_k$ stretching within ±2 eV around 400 eV and 5 values of $\theta$ stretching within ±9.5°. The electron image created in this case by HSA in its exit (focal) plane (ignoring for brevity the angular distortions caused by the hemisphere and aberrations in the HSA lens[14,15] is shown in Fig. 2 (*a*). Note that for better visibility the broadenings in this image are artificially inflated. The imaging properties of HSA ensure that the X coordinate in this image is a linear function of *E* and Y a linear function of $\theta$. It is important to take into account that the electron trajectories come into each point of the HSA exit plane focused with an inherent angular divergence in both *E*- and *θ*-directions, in our case 1.32° and 0.93° FWHM, respectively.[35] These divergent rays form the source for the subsequent iMott optics.

The electrostatic lens *EL*, following HSA, is shown in Fig. 3 (*a,b*). Operating at high accelerating voltage of 40 kV, the lens adopts a cascade design with increasing diameter of the cylindric electrodes. We note the absence of any apertures in the lens. The last electrode is connected to the Mott hemisphere forming a field-free region around the target. In our ray-tracing analysis, first, we investigated the imaging properties of this lens using an ideal 25-point source simulating the image produced by HSA, Fig. 2 (*a*). Here the spots were replaced by points and the angular divergence, as the most critical test, was increased to 4° FWHM along both directions. Fig. 3 (*a,b*) shows the electron trajectories originating from the central and two ±2 eV and ±9.5° end points of the source in EL along the two axial cross-sections in the *E*- and *θ*-directions, respectively. We note that in each point of the HSA focal plane the central trajectories are normal to this plane in the *E*-direction, but inclined in the *θ*-direction with the angle relative to the normal progressively increasing with *θ*, in our case to ±7° at the limits of the *θ*-range. The image transferred from the above point source to the Au target is shown in Fig. 3 (*c*, note the artificially inflated broadening). The lens demagnifies the image by a factor of ~3 in order to reduce aberrations in the subsequent MLs. The focusing voltages at the lens are optimized for the points slightly away from the centre. The image shows practically no distortions and is characterized by average $\Delta E$ and $\Delta \theta$ broadenings of 3.8 meV and 0.036° FWHM, respectively, which can be considered negligible compared to the HSA resolution figures even in our most critical case. Indeed, Fig. 3 (*c*) shows the image transferred from the real broadened source at the HSA focal plane, Fig. 2 (*a*). Evidently, the EL introduces practically no broadening or aberrations. As we discuss below, such extraordinary imaging properties are achieved by virtue of the high accelerating voltage. Finally, our electron optics allows efficient spin-integrated measurements by mechanical in-situ exchange of the Au target with a direct-view eCCD. As discussed below, this electron registration method much supersedes the MCP/phosphor screen/CCD stack conventionally used in ARPES analyzers.[14,35]

The image produced by electrons quasi-elastically scattered from the Au target is transferred to the eCCDs by the magnetic lens *ML1* in one scattering plane along the *E*-direction and *ML2* in the orthogonal scattering plane along the *θ*-direction, Fig. 1 (*b*). The lenses are installed at an angle of 60° relative to the incident electron beam. The use of magnetic lenses delivered in our case better performance compared to their electrostatic counterparts because their focal source and image planes are more flat. This allowed better matching of the flat extended source at the Au target to the images at the eCCDs, dramatically reducing aberrations away from the central ray. Technically, smaller size of magnetic lenses compared to electrostatic ones operating at large accelerating voltages allows compact design of our MLs restricted by the Mott hemisphere, Fig. 3. The latter is made of μ-metal, which protects HSA from leakage of magnetic fields from the lenses. The MLs and eCCDs are biased with 40 kV, equating their potential to that of the Au target. Power to each ML coil is provided by a separate 'floating point' power supply. Data transmission from the eCCDs is organized with opto-cables. The MLs are electromagnetic which allows tuning their focusing properties. Their solenoids are embraced in UHV compatible jackets allowing their air cooling through insulating plastic hoses. An aperture with a diameter of 12 mm in front of the lenses restricts the acceptance area on the target, and an iris with a diameter of 4.5 mm in the middle of the lenses restricts their angular acceptance from each point on the target to ±5°. We note that the latter, with almost isotropic distribution of the quasi-elastically scattered electrons, reduces of the geometrical lens acceptance by a factor of 6.25 in comparison with the one-channel Mott detector whose acceptance may reach ±15°.[28] Taking into account the angular dependence of the Sherman function having a lobe near the 120° scattering angle,[36] we obtain an effective reduction of the electron optics transmission by a factor of about 5.5.

Again, we started our ray-tracing simulations with the imaging properties of the magnetic lenses using the ideal 25-point source simulating the image produced at the Au target, Fig. 2 (*b*). The electron trajectories in *ML1* originating from the central point of the source are shown in Fig. 4 for the axial cross-section of the lens along the *E*-direction (*a*) and along the *θ*-direction (*b*). For clarity, the angular spread of electrons in the simulations was restricted by the *E*-direction. Importantly, although the angular spread of electrons was one-dimensional in our simulations, the resulting trajectories become three-dimensional. This is caused by the Lorentz force in the magnetic field which curls the electron trajectories. Furthermore, for the axial cross-section of *ML1* along the *E*-direction, the trajectories originating from the ±2 eV end points are shown in Fig. 4 (*c*). The best matching to the lens focal plane is achieved by inclination of the eCCDs by ~45°. The trajectories in *ML2* in the orthogonal scattering plane are identical, except that the *E*- and *θ*-directions are swapped. Images of the above 25-point point source transferred by *ML1* onto *eCCD1* and by *ML2* onto *eCCD2* are shown in Fig. 4 (*d*) and (*e*), respectively. As the magnetic field curls the electron trajectories, the rectangular pattern of the source is rotated by ~30°. The square-like vs stripe-like appearance of the *eCCD1* and *eCCD2* images expresses the fact that the view angles of *ML1* and *ML2* onto the Au target are inclined in two orthogonal planes along the *E*- and *θ*-directions, respectively. We note that the aberrations almost vanish near the central ray but significantly scale up away from the center, which is typical of electron optics working with extended sources. The rotation and distortion of the image breaks the linear and independent relations of the X-coordinate to *E* and Y-coordinate to *θ*. Their relation can nevertheless be described using an *n*-order polynomial morphing transformation defined by the equations $E = \sum_{i*j \leq n}^{3} A_{ij} X^i Y^j$ and $\theta = \sum_{i*j \leq n}^{3} B_{ij} X^i Y^j$, with the coefficients $A_{ij}$ and $B_{ij}$ determined by linear least-squares fitting. The images from in Fig. 4 (*d*) and (*e*) corrected using the $n = 3$ transformation are shown in Fig. 4 (*f*) and (*g*), respectively.

Finally, we have performed ray-tracing simulations with the real broadened source, Fig. 2 (*b*). The images analogous to those in Fig. 4 (*c,d*), rotated and distorted in the spatial coordinates, were transformed into the physical *E*- and *θ*-coordinates using the above morphing transformation. The resulting images are shown in Fig. 2 (*c,d*) for the *eCCD1* and *eCCD2*, respectively. The transformation has fully recovered the rectangular pattern of the source in the *E*- and *θ*-coordinates. With negligible contribution of the EL, the MLs introduce into these images aberrations which scale up away from the central ray and can be characterized by average <Δ*E*> and <Δ*θ*> broadenings of 50 meV and 0.2° FWHM for the *eCCD1*, and 30 meV and 0.63° for the *eCCD2*. These figures are certainly significant compared to the resolution figures of HSA itself, but can be considered acceptable in view of the efficiency gain delivered by the multichannel detection. Moreover, our ray tracing analysis suggests that the iMott resolutions can be improved by reduction of the image size on the Au foil by increasing the EL demagnification. The optimal demagnification, which will balance the aberrations of all lenses and pixel size of the eCCDs, has yet to be determined with the real instrument. In the time of writing the iMott detector described above is under construction for the soft-X-ray ARPES facility[37] at the ADRESS beamline of the Swiss Light Source.

## 4 PROPERTIES AND ADVANTAGES OF THE iMOTT CONCEPT

We start the discussion of the iMott properties with the exact definition of efficiency gain delivered by the multichannel detection. For the energy- and angle-resolving analyzer (with obvious generalization for the microscope) this figure should be defined as $G = \frac{S_{E,\vartheta}}{\langle \Delta E \rangle \langle \Delta \vartheta \rangle} \cdot \frac{T}{T_0}$, where the first term is the ratio of the intercepted area $S_{E,\theta}$ in the $(E,\theta)$ coordinates to the average resolutions product <Δ*E*><Δ*θ*> representing the single-channel detection, and the second term is the ratio of the (aperture limited) multichannel electron optics transmission *T* to the single-channel one $T_0$. In our case of the compact iMott, the above resolutions figures and aperture limited angular acceptance of the MLs yield $G = 8.4 \times 10^2$ which is almost 3 orders of magnitude.

Our ray tracing analysis shows that the aberrations dramatically increase away from the central ray, an inherent property of electron optics working with an extended source. An obvious way to further improve $\Delta E$ and $\Delta\theta$ will be to scale up the dimensions of the magnetic lenses, which will flatten the source and image focal planes and therefore reduce the aberrations. As we mentioned above, the present implementation of the iMott concept was restricted by $R_M$=100mm fitted to the PHOIBOS-150 analyzer. Larger analyzers such as PHOIBOS-225 or microscopes allow scaling up of the iMott size. Our preliminary ray tracing analysis indicates that in this case, most conservatively, the $\Delta E$ and $\Delta\theta$ resolutions improve and ML aperture increase linearly, the latter meaning quadratic increase of the ML solid angle acceptance. Correspondingly, only doubling of the iMott dimensions increases the multichannel efficiency to a colossal value of $G = 1.4 \times 10^4$.

The iMott scheme has several conceptual advantages compared to the previous multichannel spin detection schemes:[23,26]

(1) The imaging principles of the iMott electron optics imply the transfer from one image plane to another of inherently divergent electron beams, in contrast to the spin-filter ARPES analyzer[26] which relies on trade-off between collimation and focusing of the electron beams. This relieves the iMott from any need to reduce the normal beam divergence from HSA (in particular, the divergence in the $E$-direction resulting in the so-called α-factor in the HSA energy resolution[15] concomitantly reducing the electron optics transmission, or introduce any apertures to restrict the $(E,\theta)$ area delivered by HSA. The resulting high transmission of the electron optics compensates the Mott scattering FOM being much smaller than that of the low-energy spin filters;[23,26]

(2) The iMott concept positively utilizes the Liouville-Helmholtz theorem which states that the product $\Delta x \cdot \Delta \alpha \cdot \sqrt{V}$ is constant, where $\Delta x$ and $\Delta \alpha$ are the spatial and angular broadening of the electron beam, respectively, and $V$ is its energy. Simultaneously reducing $\Delta x$ and $\Delta \alpha$, the high accelerating voltages improve the focusing and thus $\Delta E_k$ and $\Delta\theta$ without compromising the electron optics transmission;

(3) The intercepted $E_k$ bandpass is limited only by HSA one rather than by the working regions of the W, Ir or $Fe_2O_3$ spin filters with an energy width varying from several eV to less than 1 eV as determined by peaks of their FOM energy dependence.[24] Furthermore, the iMott's Sherman function is practically identical for all the energies and angles which are simultaneously detected;

There are also advantages of the iMott concept on the technical and practical sides:

(1) The measurements do not require re-magnetization of the sample or detector to measure the $A(E_k,\theta)$ asymmetry inherently utilized in the spin-filter ARPES instruments. Instead, the asymmetry is derived from simultaneous measurements on two eCCDs. Furthermore, the pair of eCCDs installed in the orthogonal scattering plane simultaneously delivers the second spin component;

(2) Scattering of high-energy electrons is almost insensitive to the surface conditions of the Au target. This relieves of laborious surface preparation procedures and surface degradation problems typical of the LEED or spin-filter based detectors.

## 5. POSITION-SENSITIVE DETECTORS

Finally, we comment on the choice of position-sensitive detectors for iMott. One of the most important technical advantages of the iMott concept is that the high electron energies enable implementation of the position-sensitive detectors as directly irradiated eCCDs. Their detection efficiency with electron energies above 20 keV is nearly 100%. These devices have numerous advantages compared to the MCP/phosphor screen/CCD stacks conventionally used in ARPES for multichannel detection. Essential for the iMott concept, the eCCDs are energy selective with a bandwidth of about 200 eV which is achieved with an amplitude discriminator at their output. This allows selection of the quasi-elastically scattered electrons and rejection of the inelastically scattered ones, which carry less spin information and are focused by MLs away from the nominal imaging plane. Furthermore, the eCCDs benefit from their simplicity, larger dynamic range and possibility to measure absolute electron counts. The iMott detector can use standard, not even high-end, eCCDs such as back-thinned ones from Hamamatsu[38] which feature an active area of $12 \times 12$ mm$^2$ with a matrix of $512 \times 512$ effective pixels having a size of $24 \times 24$ μm$^2$. This delivers spatial resolution well sufficient compared to the focused spot size delivered by the iMott optics.

Maximal readout frequency of high-end eCCDs can nowadays top up 10 MHz. This ensures the absence of any charge saturation and, moreover, enables measurements in single-pulse counting mode. In this case fast online data processing to calculate the center of gravity of each event allows increase of the effective spatial resolution.

## 6. POTENTIAL APPLICATIONS

In the above, we have primarily focussed on the use of the iMott combined with the angle- and energy-resolving HSA. This configuration will be especially useful for SARPES which has so far been one of the main applications of the single-channel Mott detectors. The advantage compared to the existing experimental setups is that a spin-resolved band map will be directly obtained and, depending on the requirements, the data can be binned afterwards to enhance statistics for small signals. The colossal achieved efficiency gain will also allow for SARPES in the soft X-ray energy range which will extend from spin phenomena in the bulk such as the bulk Rashba states[5] to previously unthinkable applications to buried heterostructures, interfaces[40] and impurities[41] (see

Ref.39 for a recent review). Furthermore, SARPES can now be extended to less intense VUV- and X-ray sources including bending-magnet beamlines and laboratory sources.

Applications of the iMott spin detector are certainly not limited to its combination with HSA. Its unique capabilities can be used in combination with any instrument which produces a 2D image of electrons at its exit plane, with energies up to several keV. As described above, the high energies used in the iMott create ideal focussing conditions, and the angular divergence of the electrons delivered by the instrument it is combined with is not crucial. At most, only the first element of the electrostatic lens would have to be adapted. Based on these characteristics, we can envision the following (and not limited to) applications of the iMott. First, it can be placed behind time-of-flight analyzers such as ARTOF[16] to allow for direct spin-resolved Fermi surface mapping. Although the angular variance of electron trajectories in the MLs introduces some time-of-flight uncertainties, high electron energies render their contribution negligible. However, for this type of analyzers the eCCDs should be replaced by faster position-sensitive detectors (for example, delay-line detectors) having a readout speed better than 100 MHz to ensure sufficient energy resolution. Second, the iMott can be combined with a photoemission electron microscope (PEEM) for imaging of magnetic domains even with unpolarised light. This could open the possibility of lab based measurements to complement the current synchrotron based experiments. Furthermore, one could use the PEEM in combination with circular polarised light for spatial resolved chiral imaging of molecules.[42] Third, combined with a low energy electron microscope (LEEM) the iMott can be used for magnetic imaging, or for the direct visualisation of spin transfer torque through a thin layer when combined with a polarised source. Even more ambitious would be the first direct visualisation of spatial entanglement in a solid.

In connection with the completion and planned construction worldwide of several X-ray FEL facilities, delivering short and very intense X-ray pulses, we note that in contrast to other spin detection schemes the iMott is perfectly suited for time-resolved studies, especially under influence of a so-called jitter of the FEL pulses. This virtue comes because the spin-asymmetry is obtained in a simultaneous measurement which does not have to be repeated with different FEL pulses and different sample conditions. For each pulse, the complete data acquired at every eCCD can be stored and later analysed accordingly, as is now common practice for spin-integrated measurements at such facilities,[43] in order to recover the whole time evolution picture. The readout frequency of the eCCDs up to 10 MHz enables seamless handling of each pulse even at the highest repetition rates, which is presently 27 kHz at the European XFEL. We will now comment on possible saturation of the eCCDs with the very intense FEL pulses. Recent spin-polarized photoemission experiments at FLASH[44] with a single-channel Mott detector have found that more than one electron per pulse to reach the detector of passivated implanted planar silicon (PIPS) which was prohibitive for single-pulse counting. However, these experiments were essentially angle- and energy-integrated, bringing to one detector the whole integral photoemission intensity. On the other hand, the most recent time-resolved HAXPES experiments at SACLA[45] have shown that every pulse produces about $3\times10^6$ photoelectrons. For the actual iMott design, with scattering at the Au target characterized by an efficiency of the order of $10^{-1}$ of the total (inelastic and elastic) reflectivity and with the actual iris of the MLs, only about $5\times10^4$ electrons will pass to the eCCDs. They will distribute over more than $2.5\times10^5$ channels of our eCCD, leaving about one-fifth electron per channel. This number of events remains sufficiently low for single-pulse counting and thus energy resolution of the scattered electrons essential for the iMott operation. However, given the very approximate character of this estimate, we cannot rule out that certain pulse energy attenuation may still be necessary. In any case, the problem of single-pulse counting in iMott is less restrictive compared to that of space charge,[44] the main encumbrance of photoelectron spectroscopy at FEL sources, which is relieved by increase of the pulse repetition rate with simultaneous reduction of intensity of each pulse.

## 7. CONCLUSION

We have presented a concept of the multichannel electron spin detector iMott which combines imaging electron optics principles to achieve multichannel detection with the Mott scattering as the spin selective process. The detector can be fitted to a standard (photo) electron analyzer to yield a spin-resolved image in the energy and angle coordinates, or microscope to yield image in the spatial coordinates. The iMott electron optics uses consecutive imaging principles: (1) The multichannel electron image from the analyzer or microscope focal plane is re-imaged by the electrostatic lens at an accelerating voltage of 40 kV onto the Au target; (2) Quasi-elastic electrons bearing spin asymmetry of the Mott scattering are imaged by four magnetic lenses in two orthogonal planes onto energy-selective position-sensitive eCCDs to yield the multichannel spin asymmetry image. Ray-tracing calculations for a compact detector with $R_M = 100$ mm fitted to a standard HSA have demonstrates an efficiency gain of $8.4\times10^2$ compared to the single-channel detector. Scaling up of the iMott dimensions dramatically improves the resolutions and electron optics transmission, pushing the gain to a colossal factor above $10^4$ which becomes $1.4\times10^4$ already for $R_M = 200$ mm. The iMott concept has a few fundamental advantages: (1) The imaging electron optics accepts divergent electron sources typical of the electron analyzers or microscopes; (2) High accelerating voltage ensures almost ideal imaging properties of the EL stage, with replacement of the Au target by an eCCD allowing efficient spin-integrated measurements; (3) The use of directly irradiated eCCDs with their simplicity, large dynamic range and fast readout; (4) Stability and simultaneous measurements of two spin components inherited from the Mott detectors. By virtue of the colossal efficiency gain, the iMott concept enables expansion of the spin-resolved spectroscopies from standard bulk and surface systems (Rashba effect, topological insulators, magnetic pairing in unconventional superconductors, etc.) to previously unthinkable cases to buried heterostructures actual for the nowadays device applications. Furthermore, the simultaneous spin detection and fast eCCDs readout enable efficient exploitation of the iMott detectors at not only synchrotron but also X-ray FEL facilities.

*Acknowledgements*. The authors thank J. Friso van der Veen for attracting their attention to the challenge of multichannel spin detection and for promoting discussions. We thank G. Schönhense for encouragement and fruitful scientific exchange. Logistic support of the iMott project by F. Nolting and T. Schmitt is gratefully acknowledged. We are obliged to S. Maehl for providing ray-tracing data of the PHOIBOS-150 analyzer, and S. Muff for the CAD drawings. The project is partially supported by the Swiss National Science Foundation.

# FIGURE CAPTIONS

Fig. 1. (*a*) Schematics the iMott detector attached to the standard angle- and energy resolving HSA; (*b*) Blow-up of the iMott, which includes the imaging electron lens *EL*, Au target, four magnetic lenses *ML*s and position-sensitive detectors *eCCD*s. Differential images between the opposite eCCDs yield a multichannel image of the spin asymmetry $A(E,\theta)$. The use of imaging principles allows iMott to work with divergent electron beams.

Fig. 2. Ray-tracing simulations of the iMott electron optics. (*a*) The HSA focal plane. This is the source for the iMott optics, with the electron trajectories in each point being inherently divergent in both *E*- and $\theta$-directions; (*b*) The Au target. For better visibility, the broadening in (*a*) and (*b*) is shown artificially inflated by ×2 in both directions; (*c,d*) The position-sensitive *eCCD1* (and the opposite *eCCD3*) and *eCCD2* (*eCCD4*). The images are rendered into the *E*- and $\theta$-coordinates using *n*=3 polynomial morphing transformation.

Fig. 3. Imaging properties of the electrostatic lens *EL*: (*a,b*) Lens schematics and electron trajectories in the two axial cross-section along the *E*- and $\theta$-directions, respectively, originating from the central and two ±2 eV and ±9.5° end points; (*c*) Image at the Au target formed by the ideal 25-point source. The broadening is shown artificially ×10 inflated. Due to high accelerating voltage *EL* introduces negligible aberrations.

Fig. 4. Imaging properties of the magnetic lenses *ML*: (*a,b*) Lens schematics and electron trajectories in *ML1* (and opposite *ML3*) lens for the axial cross-sections of the lens in the *E*- and $\theta$-directions, respectively, originating from the central point at the Au target with angular spread in the *E*-direction; (*c*) Trajectories for the axial cross-section of *ML1* (*ML3*) along the *E*-direction originating from the ±2 eV end points. The trajectories in the *ML2* (*ML4*) plane are identical but the *E*- and $\theta$-directions are swapped; (*d,e*) Images at the *eCCD1* and *eCCD2* formed from the ideal 25-point source of quasi-elastically scattered electrons at the Au target by *ML1* and *ML2*, respectively, with their view angles inclined relative to the source in two orthogonal planes. The images are rotated because the magnetic field curls the electron trajectories; (*f,g*) The images (*d*) and (*e*), respectively, rendered into the *E*- and $\theta$-coordinates using *n*=3 polynomial morphing transformation.

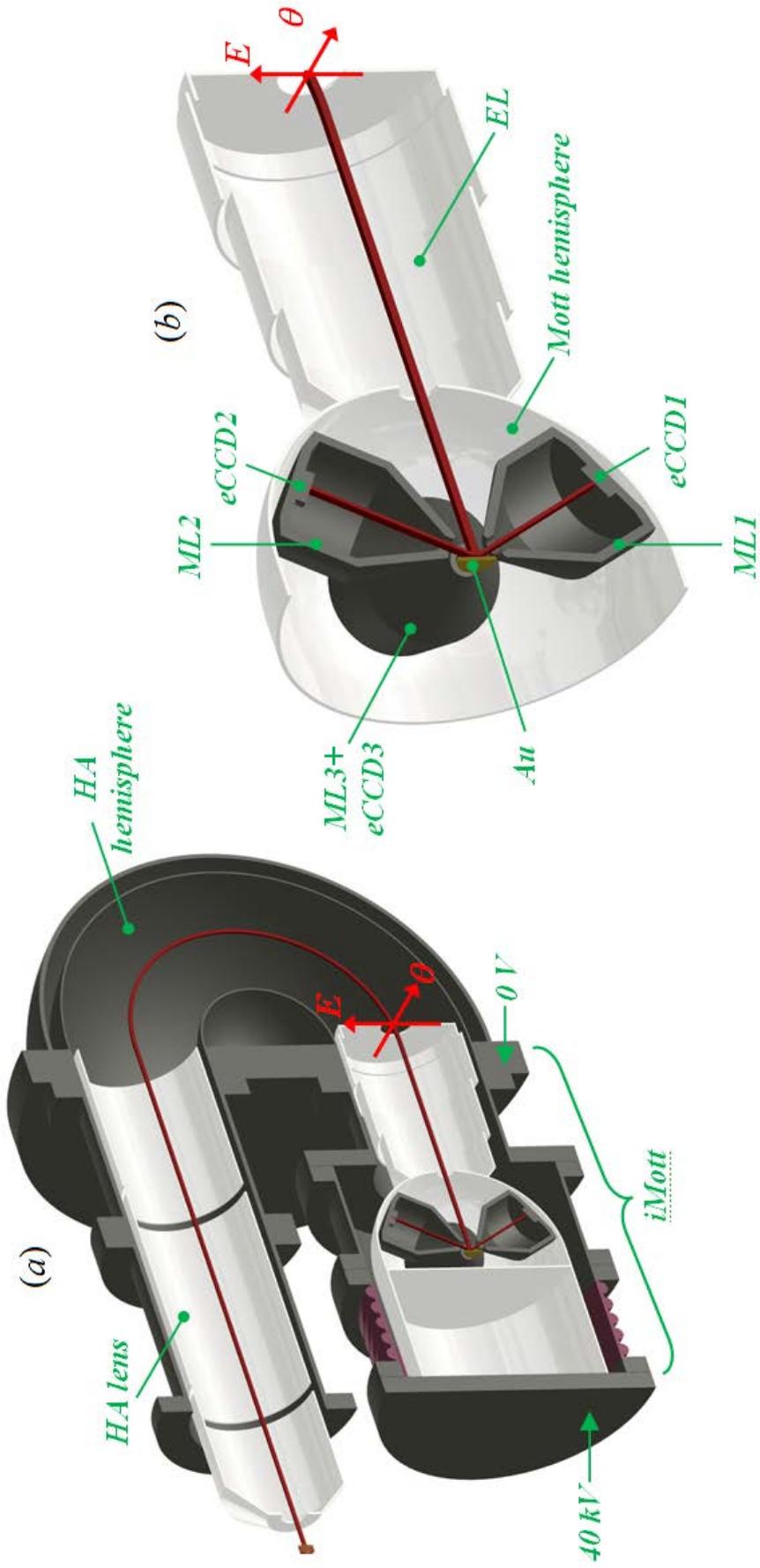

Fig. 1

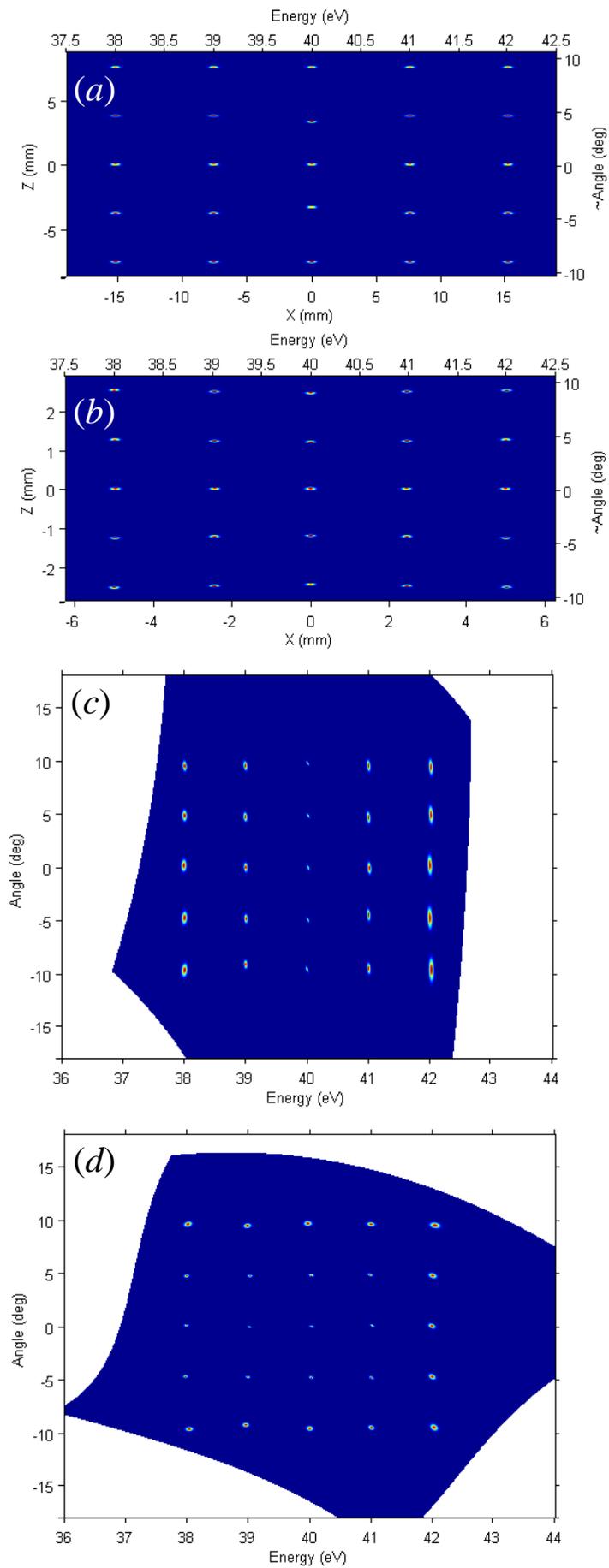

Fig. 2

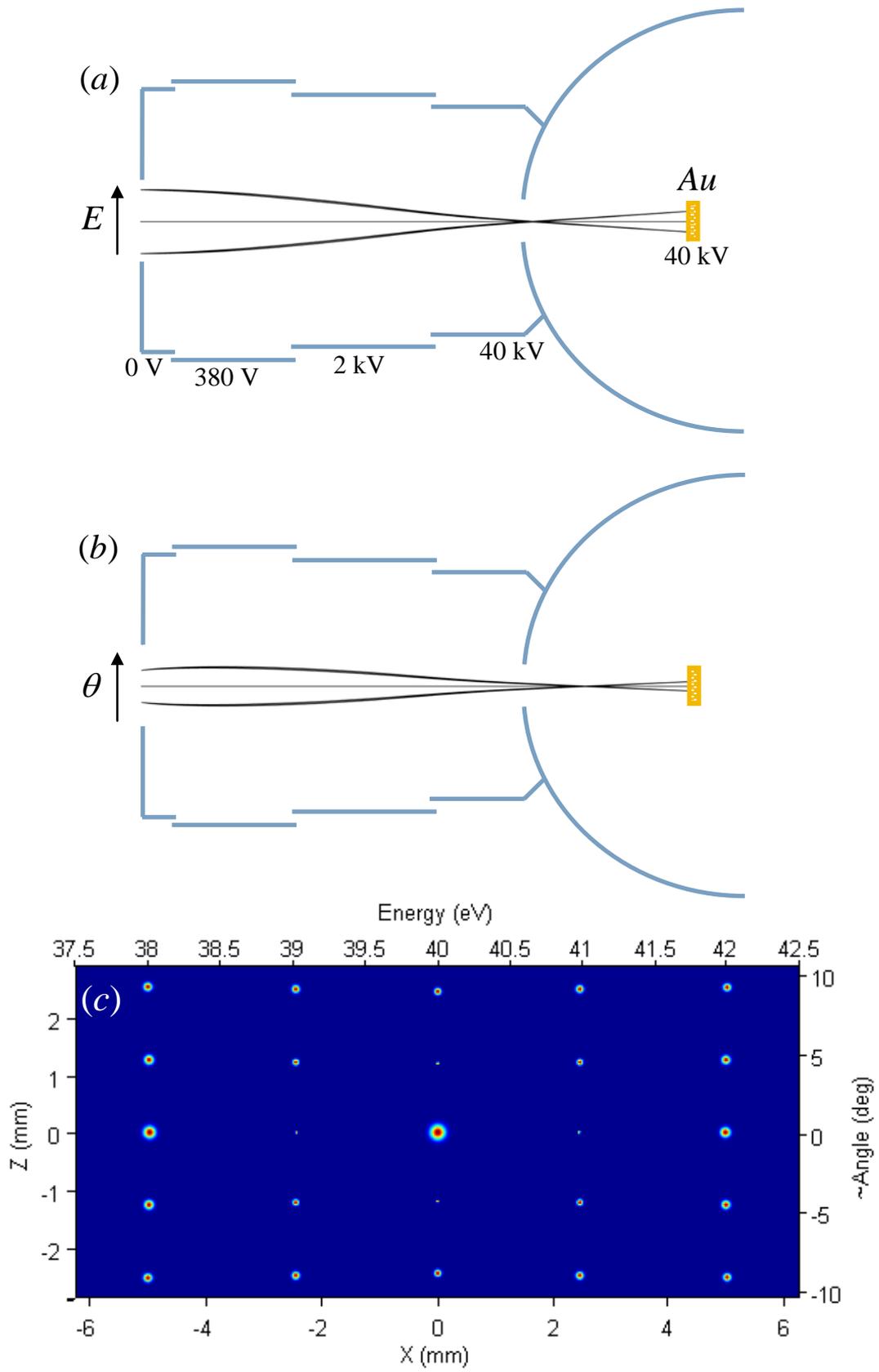

Fig. 3

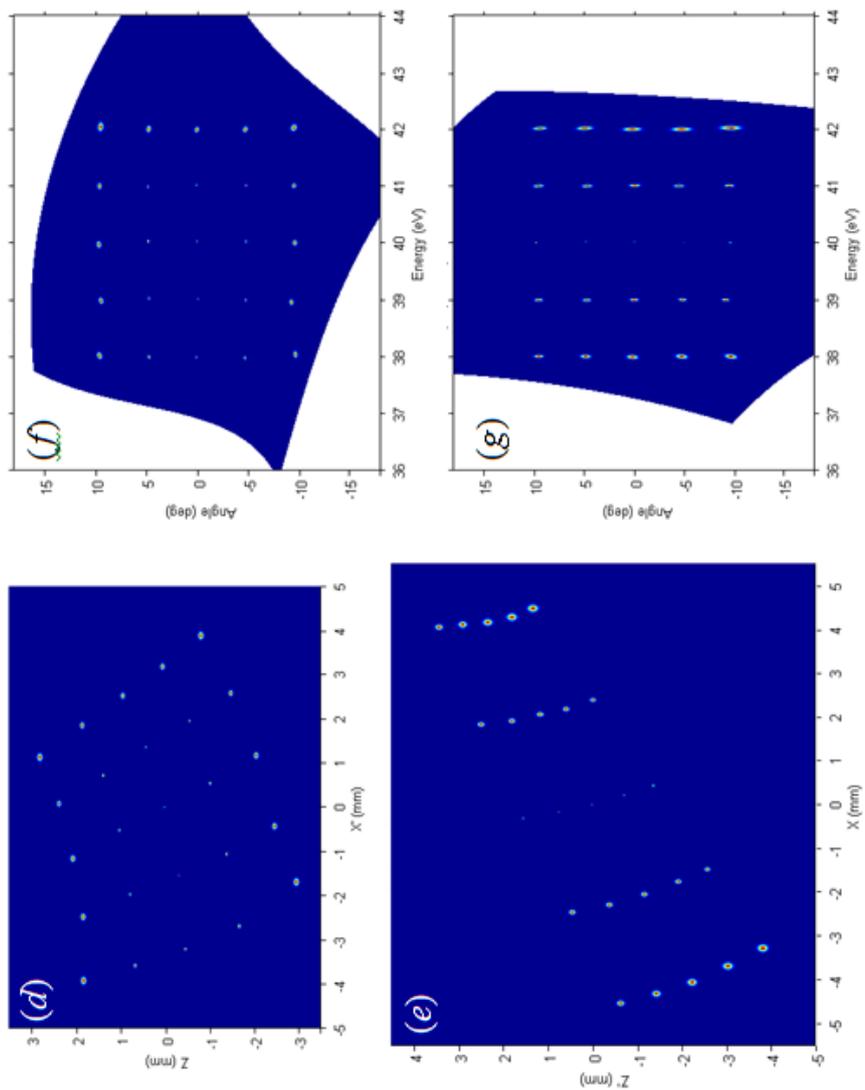

Fig. 4